\documentclass{article}
\usepackage{spconf,amsmath,graphicx}
\usepackage{graphicx}
\usepackage{color}

\usepackage{multirow}
\usepackage{lipsum}  
\usepackage{booktabs,siunitx}
\usepackage{colortbl}
\usepackage{amssymb}
\usepackage{pifont}
\usepackage{amsfonts}
\usepackage{tabularx}
\usepackage{array, makecell}
\usepackage{xcolor}
\usepackage{graphicx}
\usepackage{kotex}
\usepackage{comment}
\usepackage{soul}
\usepackage{url}

\title{Toward universal text-to-music retrieval}
\name{$^\flat$SeungHeon Doh, $^\natural$Minz Won, $^\sharp$Keunwoo Choi, $^\flat$Juhan Nam} 
\address{$^\flat$Graduate School of Culture Technology, KAIST, South Korea \\ $^\natural$ByteDance, USA \\ $^\sharp$Gaudio Lab, South Korea
}

\begin{document}
\ninept
\maketitle
\begin{abstract}
This paper introduces effective design choices for text-to-music retrieval systems. An ideal text-based retrieval system would support various input queries such as pre-defined tags, unseen tags, and sentence-level descriptions. In reality, most previous works mainly focused on a single query type (tag or sentence) which may not generalize to another input type. Hence, we review recent text-based music retrieval systems using our proposed benchmark in two main aspects: input text representation and training objectives. Our findings enable a universal text-to-music retrieval system that achieves comparable retrieval performances in both tag- and sentence-level inputs. Furthermore, the proposed multimodal representation generalizes to 9 different downstream music classification tasks. We present the code and demo online.\footnote{https://seungheondoh.github.io/text-music-representation-demo/}


\end{abstract}

\section{Introduction}
\label{sec:intro}

The demand for efficient music retrieval has been increasing as massive music libraries become easily accessible. While various methods have been proposed for efficient retrieval \cite{ghias1995query, muller2007lyrics, watanabe2019query, lee2020metric}, text-based\footnote{Because the terminology-\textbf{\{text, language\}} encompasses various input lengths, we explicitly distinguish between the tag-level and sentence-level.} retrieval remains the most prevalent \cite{turnbull2008semantic, nam2018deep, won2020evaluation}. Text-based retrieval is challenging because it needs to handle not only editorial metadata (e.g., title, artist, release year) but also semantic information (e.g., genre, mood, theme). Furthermore, modern retrieval systems, such as voice assistants \cite{sciuto2018hey}, need to generalize to sentence-level natural language inputs beyond fixed tag vocabularies.


While much research has addressed text-based retrieval, there are two dominant approaches: classification and metric learning. Classification models~\cite{choi2016automatic, nam2018deep} are trained with a set of fixed tag labels, and then the predicted tags are utilized in retrieval. Despite its successful classification performance, this approach is limited to a fixed vocabulary. In contrast, metric learning models are more flexible by using pre-trained word embeddings~\cite{choi2019zero,won2020multimodal} or language models~\cite{won2021emotion, chen2022learning, huang2022mulan, manco2022contrastive}. Especially pre-trained language models enable free-form text inputs for music retrieval by representing sentence-level semantics. There are multiple loss functions (e.g., triplet loss, contrastive loss) for metric learning based on its training objective.

\begin{figure}[!t]
\centering
\includegraphics[width=\columnwidth]{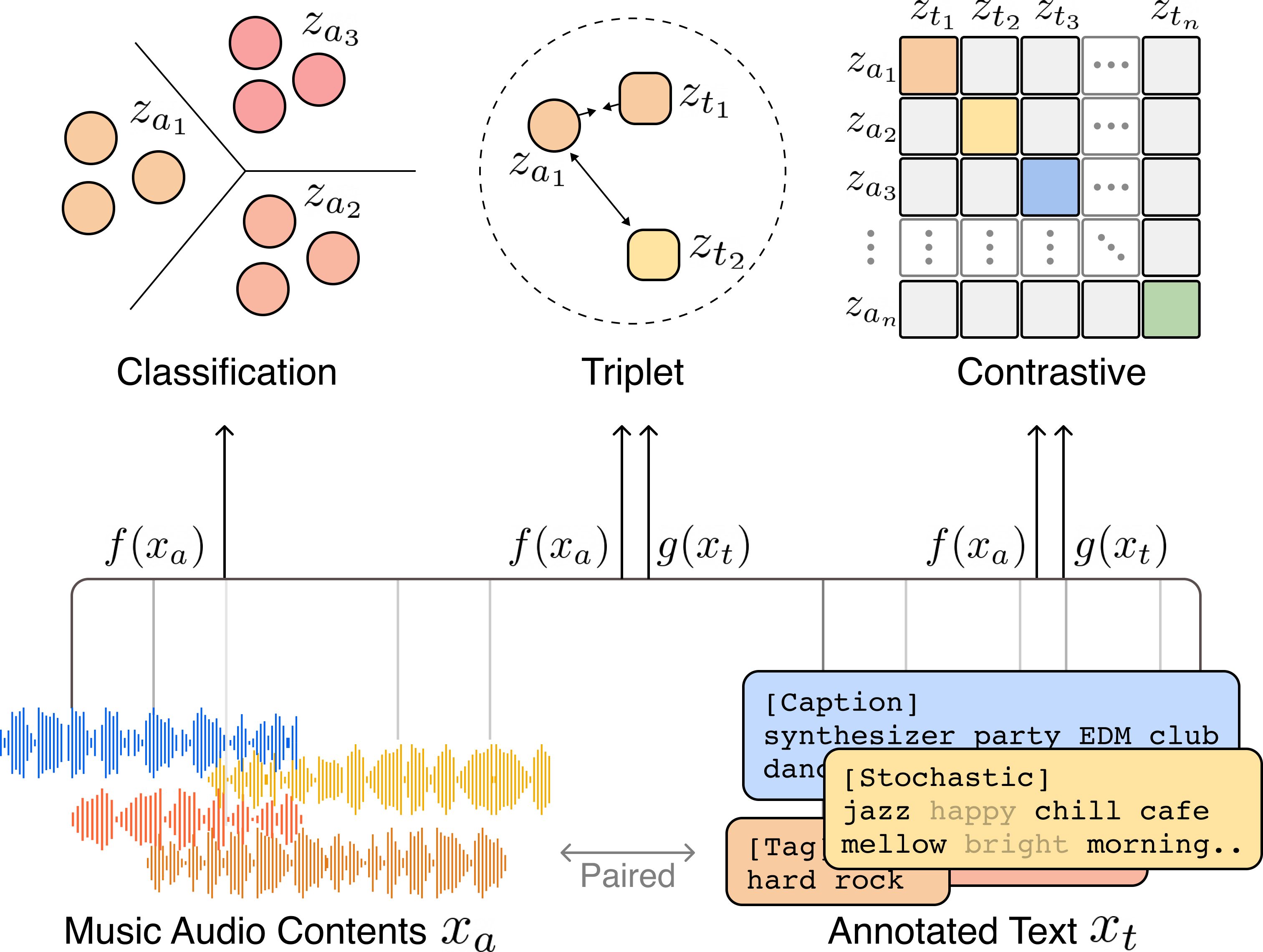}
\caption{Text-Music Representation Learning Models.}
\label{fig:main}
\end{figure}

An ideal text-based retrieval system needs to be flexible to allow various input types (e.g. word, sentence) and abundant vocabularies. For example, one can use broadly used tags, such as genre, to explore the music library. Sometimes the input queries may include unseen types of music tags. Also, another can use more detailed sentence-level descriptions to discover music. However, to the best of our knowledge, previous works mainly focused on improving a single type of input queries. Also, they are using respective datasets and evaluation metrics which makes it difficult to choose the appropriate solution for universal music retrieval.



To address this issue, we perform holistic evaluation of recently proposed text-to-music retrieval approaches. First, we review the training objectives and modality encoding of the previous works. We also propose a novel stochastic sampling of text inputs to enable a generalizable text encoder (Section 2). We then introduce a text-music paired dataset and an evaluation benchmark to assess the system's generalizability (Section 3). Section 4 depicts experimental results. Finally, Section 4 and 5 propose reusable insights for designing universal text-to-music retrieval systems.

\begin{table*}[!t]
\resizebox{\textwidth}{!}{
\begin{tabular}{lcccccccccccccc}
\toprule
MSD Subset & \# of Track & \# of Artist & \# of Album & \# of Tag & \# of Caption & Avg.Tag & A/S & Genre & Style & Inst. & Vocal & Mood & Theme & Culture \\ \midrule
Top50s~\cite{choi2016automatic, nam2018deep} & 241,889$^{\color{white}*}$  & 25,239 & 67,495 & 50 & 11,418 & 1.72 & No & $$\checkmark$$ &  & $\checkmark$ & $\checkmark$ & $\checkmark$ &  & $\checkmark$ \\
CALS~\cite{won2021transformer} & 233,147$^{*}$  & 24,569 & 63,349 & 50 & 4,408 & 1.31 & Yes & $\checkmark$ &  & $\checkmark$ &  & $\checkmark$ &  & $\checkmark$ \\
ECALS (\textbf{Ours}) & 517,022$^{\color{white}*}$  & 32,650 & 89,920 & 1054 & 139,541 & 10.18 & Yes & $\checkmark$ & $\checkmark$ & $\checkmark$ & $\checkmark$ & $\checkmark$ & $\checkmark$ & $\checkmark$ \\ \bottomrule
\end{tabular}
}
\caption{Comparison of the existing MSD-subset and the proposed ECALS subset. A/S stands for artist stratified. (*) CALS includes additional un-annotated tracks for semi-supervised learning. This table only shows tag annotated dataset.}
\label{tab:msd_subset}
\end{table*}

\section{Music and Text Representation Learning}\label{sec:method}
This section introduces recent text-music representation learning models (illustrated in Figure~\ref{fig:main}). We carefully review three training objectives and their modality encoders. In the following descriptions, $x^{a}$ denotes a music audio example, $x^{t}$ its paired text data, $f(\cdot)$ an audio encoder, and $g(\cdot)$ a text encoder. Each modality input is processed by the corresponding encoder $f$ or $g$ (described in \ref{encoding}). Each encoder consists of a backbone model, a linear projection, and an $l_{2}$ normalization layer. We denote the two output embeddings of audio and text as $z^{a} = f(x^{a})$ and $z^{t} = g(x^{t})$, respectively. The classification model does not have a text encoder since we directly perform multi-label classification of the tags.

\subsection{Training Objective}\label{sec:method:frameworks}
\vspace{2mm}
\noindent \textbf{Classification Model} The goal of the classification model is to learn a linearly discriminate embedding space. This can also be interpreted from the similarity-based metric learning perspective as introduced in \cite{lee2020metric}. The prediction score of the model for each class is $\hat{y} = sigmoid(z^{a} \cdot c_{y})$, where $c_{y}$ is a centroid vector for each class (parameters of the last dense layer). To maximize the similarity between $z^a$ and $c_y$, the objective function is formulated as follows.

\begin{equation}
\mathcal{L}_{ce} = -{(y_{z}\log(\hat{y}) + (1 - y_{z})\log(1 - \hat{y}))}
\end{equation}
Since the prediction score of a track is utilized as a similarity score with the centroid vector (the tag label), the classification-based model serves as the baseline system for tag-based retrieval. The classification model is limited to a fixed vocabulary since it cannot take advantage of text embeddings in a zero-shot retrieval scenario.


\vspace{3mm}
\noindent \textbf{Triplet-Loss Model} The goal of triplet-loss models is to learn an embedding space where relevant input pairs are mapped closer than irrelevant pairs in the latent space. The objective function is formulated as follows:

\begin{equation}
\mathcal{L}_{a \rightarrow t}=[0, \delta - z^{a} \cdot z^{t}_{pos} + z^{a} \cdot z^{t}_{neg}]_{+}
\end{equation}
where $\delta$ is the margin, $z^{t}_{pos}$ denotes the paired text for the music audio, and $z^{t}_{pos}$ denotes irrelevant text. [·]+ indicates a rectified linear unit. In practice, an efficient negative sampling is crucial in triplet-based metric learning. We applied the distance-weighted sampling method used in \cite{won2020multimodal}. Using structure-preserving constraints~\cite{wang2016learning,kim2021learning}, we utilize a symmetric loss function: $\mathcal{L}_{a \leftrightarrow t} = (\mathcal{L}_{a \rightarrow t} + \mathcal{L}_{t \rightarrow a})/2$.

\vspace{3mm}
\noindent \textbf{Contrastive-Loss Model}
The core idea of contrastive-loss models is to reduce the distance between positive sample pairs while increasing the distance between negative sample pairs. Unlike triplet-loss models, contrastive-loss models can utilize a large number of negative samples that exist in a mini batch $N$. During training, the audio and text encoders are jointly trained to maximize the similarity between $N$ positive pairs of (music, text) associations while minimizing the similarity for $N \times (N-1)$ negative pairs. This is known as the multi-modal version of InfoNCE loss \cite{oord2018representation, radford2021learning} and formulated as follows:

\begin{equation}
\label{eq2}
\begin{aligned}
\mathcal{L}_{a \rightarrow t} = - \log \frac{\exp(z^{a}_{i} \cdot z^{t}_{i} / \tau)}{\sum^{N}_{j=1} \exp( z^{a}_{i} \cdot z^{t}_{j}) / \tau)} \\
\end{aligned}
\end{equation}
where $\tau$ is a learnable parameter. The loss function is designed as follows: $\mathcal{L}_{a \leftrightarrow t} = (\mathcal{L}_{a \rightarrow t} + \mathcal{L}_{t \rightarrow a})/2
$.


\subsection{Audio Encoding}
\label{encoding}
For all experiments, we utilize a modified version of Music Tagging Transformer~\cite{won2021transformer} as our audio encoder. The first four convolution layers capture local acoustic features and the following four transformer layers summarize the sequence. The output of convolutional layers (audio sequence) attaches \textsc{[CLS]} token at the first position, and the output of the last layer of the transformer at the \textsc{[CLS]} token is treated as the feature that represents the whole audio. It is, finally, linearly projected into an embedding space. We use mel spectrograms as input without any augmentation. 

\begin{table*}[!t]
\centering
\resizebox{0.8 \textwidth}{!}{
\begin{tabular}{llllcclccccc}
\toprule
\multicolumn{4}{l}{} & \multicolumn{2}{c}{Tag-level Retrieval} &  & \multicolumn{5}{c}{Sentence-level Retrieval} \\  \cmidrule{5-6} \cmidrule{8-12} 
\multirow{2}{*}{Model Type} & \multirow{2}{*}{Text Enc.} & \multirow{2}{*}{Text Rep.} & \multirow{2}{*}{Used In} & 50 Tags & 1054 Tags &  & \multicolumn{5}{c}{1000 Captions} \\
 &  &  &  & ROC/PR & ROC/PR &  & R@1 & R@5 & R@10 & mAP10 & MedR$\downarrow$ \\ \midrule
Classification & Binary & Tag & \cite{choi2016automatic, won2021transformer, lee2020metric} & 90.2 / 39.5 & \textbf{86.4} / \textbf{8.8} &  & 4.0 & 13.8 & 20.1 & 8.3 & 86 \\
Triplet & GloVe & Tag & \cite{choi2019zero,won2020multimodal} & 89.2 / 36.0 & 82.6 / 6.1 & & 2.8 & 11.2 & 18.6 & 6.6 & 51.5 \\
Triplet & GloVe & Sentence & Ours & 88.6 / 37.1 & 76.8 / 5.3 &  & 5.4 & 22.1 & 35.0 & 13.0 & 17 \\
Triplet & GloVe & Stochastic & Ours & 89.2 / 37.6 & 81.6 / 6.2 &  & 6.4 & 21.8 & 32.7 & 12.8 & 19.5 \\
Triplet & BERT$_{base}$ & Tag & Ours & 86.9 / 30.2 & 81.7 / 5.1 &  & 1.6 & 6.2 & 12.0 & 3.9 & 68 \\
Triplet & BERT$_{base}$ & Sentence & \cite{won2021emotion} & 87.7 / 35.0 & 78.8 / 5.4 &  & 6.7 & 23.6 & 36.6 & 14.1 & 16 \\
Triplet & BERT$_{base}$ & Stochastic & Ours & 88.4 / 35.0 & 83.6 / 6.3 &  & 6.6 & 25.1 & 39.4 & 14.6 & 16 \\
Contrastive & BERT$_{base}$ & Tag & Ours & \textbf{90.6} / \textbf{40.2} & \textbf{86.4} / \textbf{8.8} &  & 2.5 & 13.7 & 22.5 & 7.4 & 47 \\
Contrastive & BERT$_{base}$ & Sentence & \cite{manco2022contrastive,huang2022mulan} & 87.0 / 32.5 & 77.6 / 5.1 &  & 6.8 & 25.4 & 38.4 & 15.3 & 17 \\
Contrastive & BERT$_{base}$ & Stochastic & Ours & 89.8 / 38.0 & 84.8 / 7.7 &  & \textbf{10.2} & \textbf{29.8} & \textbf{42.8} & \textbf{18.7} & \textbf{13} \\ 
\bottomrule
\end{tabular}
}
\caption{Tag based, and Sentence based Retrieval result. \textbf{Used In} refers to previous studies using the same method.}
\label{tab:text_retireval}
\end{table*}

\subsection{Text Encoding}
We use tag and sentence text representation for input of the text encoders. For this, we use a pre-trained word embedding GloVe~\cite{pennington2014glove} and a pre-trained Bidirectional Encoder Transformer (BERT)~\cite{devlin2018bert} with a base-uncased architecture. In using both text encoders, tag and sentence representations are processed differently. The tag representation uniformly samples one tag among multi-label texts. The sentence representation uses the entire multi-label text by concatenating multi-label text. In the case of the GloVe model, the sentence text is tokenized by white space, and projected to joint embedding space, then average the sentence embedding sequence \footnote{We tested early fusion and late fusion of word embedding, but there was no significant difference in the results.}. In the case of BERT model, the input text sequence is tokenized by wordpiece tokenizer, and the max sequence length is 64. Similar to audio feature embedding, the text sequence attaches \textsc{[SOS]} token at first position and the output of the last layer of the transformer at the \textsc{[SOS]} token are treated as the feature representation of the text which is layer normalized and then linearly projected embedding space.

\subsection{Stochastic Text Representation}
In the preliminary study, we find that there is a strong association between text representation (train stage) and text query types (test stage). As somewhat obviously, the model works better when the input forms during the training phase and test phase are homogeneous, there are no references studying the relationship between text representation and retrieval performance. To use the advantages of both, we propose a stochastic text representation. During the training stage, we select $K$ words from $L$ length text sentence. At this time, $K$ is uniformly randomly sampled among integer numbers from $1$ (word length) to $L$ (sentence length). Unlike the dropout method, which determines the length by probability value, stochastic sampling has a dynamic input length.

\section{EXPERIMENT}
\label{sec:exp}
\subsection{Music-Text Pair Dataset (ECALS) 
\label{sec:ecals}}
With the growing interest in sentence-level retrieval tasks~\cite{huang2022mulan,manco2022contrastive}, it is desirable to have a music-caption paired dataset. However, no dataset is available for re-implementation. To address this problem, we concatenate the tag from different annotation sources. Based on Million Song Dataset (MSD)~\cite{bertin2011million}, we propose the ECALS (Extended Clean tag and Artist-Level Stratified) subset by merging the CALS subset \cite{won2021transformer} with 500 Last.fm tags \cite{won2020multimodal} and 1,402 AllMusic \cite{schindler2019multi} tag annotation. As a result, the ECALS subset has 0.52 million 30-second clips and 140k unique tag captions, including genre, style, instrument/vocal, mood/theme, and culture categories. Table \ref{tab:msd_subset} shows the size and statistics of the MSD subset. The test track of the ECALS subset is the same as the CALS subset, and only the train, validation track, and annotation tags have been increased. Using the ECALS dataset, we evaluate tag-level and sentence-level retrieval tasks.


\begin{table}[!t]
\centering
\resizebox{\columnwidth}{!}{%
\begin{tabular}{lccccc}
\toprule
Dataset & Task & \# of Track & \# of Tag & Avg.Tag & Metric \\ \midrule
MTAT & Tagging & 25,860 & 50 & 2.70 & ROC/PR \\
MTG-top50s & Tagging & 54,380 & 50 & 3.07  & ROC/PR \\
MTG-G & Genre & 55,000 & 87 & 2.44  & ROC/PR \\
FMA-Small & Genre & 8,000 & 8 & 1.00  & ACC \\
GTZAN & Genre & 930 & 10 & 1.00  & ACC \\
MTG-I & Instrument & 24,976 & 40 & 2.57  & ROC/PR \\
KVT & Vocal & 6,787 & 42 & 22.78  & F1 \\
MTG-MT & Mood/Theme & 17,982 & 56 & 1.77  & ROC/PR \\
Emotify & Mood & 400 & 9 & 1.00  & ACC \\ \bottomrule
\end{tabular}
}
\caption{Downstream tasks/datasets for music semantic}
\label{tab:downstream}
\end{table}

\subsection{Evaluation Dataset}
For unseen-query retrieval and downstream evaluation, we select various datasets related to music semantic understanding. The selection criteria are as follows: if a dataset has 1) commercial music for retrieval, 2) publicly assessed (at least upon request) and 3) categorical single or multi-label annotations for supporting text-based retrieval scenarios. We summarize all the datasets and tasks in Table \ref{tab:downstream}. MagnaTagATune (MTAT) \cite{law2009evaluation} consists of 26k music clips from 5,223 unique songs. Following a previous work \cite{castellon2021codified, huang2022mulan}, we use their published splits and top~50 tags. We do not compare the result with previous works using different split \cite{manco2022contrastive, manco2022learning}. MTG-Jamendo (MTG) \cite{bogdanov2019mtg} contains 55,000 full audio tracks with 195 tags about genre, instrument, and mood/theme. We use the official splits (\textit{split-0}) in each category for tagging, genre, instrument, and mood/theme tasks. For single-label genre classification, we use the fault-filtered version of GTZAN (GZ) \cite{sturm2013gtzan} and the `small' version of Free Music Archive \cite{defferrard2016fma} (FMA-Small). For the vocal attribute recognition task, we use K-pop Vocal Tag (KVT) dataset \cite{kim2020semantic}. It consists of 6,787 vocal segments from K-pop music tracks. All the segments are annotated with 42 semantic tags describing various vocal styles including pitch range, timbre, playing techniques, and gender. For the categorical mood recognition task, we use the Emotify dataset \cite{aljanaki2016studying}. It consists of 400 excerpts in 4 genres with 9 emotional categories.

\begin{table*}[!t]
\centering
\resizebox{\textwidth}{!}{%
\begin{tabular}{lllcccccccccccc}
\toprule
 &  &  & \multicolumn{2}{c}{Tagging} &  & \multicolumn{3}{c}{Genre} &  & \multicolumn{2}{c}{Mood/Theme} &  & \multicolumn{2}{c}{Instrument/Vocal} \\
 &  &  & MTAT & MTG-Top50s &  & MTG-G & GZ & FMA &  & MTG-MT & Emot &  & MTG-I & KVT \\ 
Model Type & Text Enc. & Text Rep. & ROC/PR & ROC/PR &  & ROC/PR & ACC & ACC &  & ROC/PR & ACC &  & ROC/PR & F1 \\ \midrule
\multicolumn{15}{l}{{\color[HTML]{656565} \textit{Zero-shot Transfer:}}} \\
Classification & Binary & Tag & - & - &  & - & - & - &  & - &  &  & - & - \\
Triplet & GloVe & Tag & 75.45 / 19.77 & 74.21 / 22.34 &  & 80.42 / 14.52 & 86.21 & 44.88 &  & 63.58 / 6.42 & 21.25 &  & 55.85 / 9.04 & 68.96 \\
Triplet & GloVe & Sentence & 72.08 / 18.24 & 73.55 / 22.89 &  & 79.79 / 15.14 & 86.90 & \textbf{48.12} &  & 61.28 / 5.87 & 11.25 &  & 55.50 / 9.16 & 68.98 \\
Triplet & GloVe & Stochastic & 72.96 / 18.39 & 74.51 / 22.74 &  & 81.17 / 14.93 & 87.24 & 45.75 &  & 63.36 / 6.79 & 10.00 &  & 54.95 / 9.05 & 69.03 \\
Triplet & BERT$_{base}$ & Tag & 74.41 / 17.60 & 74.34 / 20.91 &  & 79.87 / 13.03 & 77.59 & 39.00 &  & 63.69 / 6.88 & 21.25 &  & 54.37 / 9.20 & 69.14 \\
Triplet & BERT$_{base}$ & Sentence & 73.21 / 18.82 & 75.69 / 22.83 &  & 81.55 / 14.97 & 85.86 & 39.38 &  & 59.65 / 6.59 & 15.00 &  & 57.73 / 9.44 & 69.96 \\
Triplet & BERT$_{base}$ & Stochastic & 74.83 / 19.85 & 75.67 / 23.10 &  & 80.66 / 14.89 & 87.24 & 41.38 &  & 65.88 / 7.70 & 27.50 &  & 59.84 / 10.38 & 69.98 \\
Contrastive & BERT$_{base}$ & Tag & 77.34 / \textbf{21.96} & \textbf{76.39} / \textbf{24.48} &  & \textbf{81.76} / \textbf{16.85} & \textbf{89.31} & 47.38 &  & \textbf{66.86} / \textbf{8.67} & 17.50 &  & 60.95 / \textbf{11.40} & \textbf{70.43} \\
Contrastive & BERT$_{base}$ & Sentence & 74.22 / 19.49 & 75.56 / 21.55 &  & 81.56 / 14.39 & 78.97 & 39.38 &  & 62.32 / 6.52 & 18.75 &  & \textbf{61.40} / 10.20 & 69.95 \\
Contrastive & BERT$_{base}$ & Stochastic & \textbf{78.41} / 21.23 & 76.14 / 23.60 &  & 81.19 / 15.57 & 87.93 & 45.12 &  & 65.66 / 8.09 & \textbf{33.75} &  & 60.64 / 11.26 & 70.35 \\ 
\rowcolor[HTML]{EFEFEF}
\multicolumn{3}{l}{\color[HTML]{656565}State-of-the-art \cite{huang2022mulan,choi2019transfer}} & \color[HTML]{656565}{78.2 / -} & - &  & - & \color[HTML]{656565}{73.1} & - &  & - & - &  & - & - \\ \midrule
\multicolumn{15}{l}{{\color[HTML]{656565}\textit{Probing:} }} \\ 
Classification & Binary & Tag & 89.72 / 35.54 & 82.66 / 28.78 &  & 87.01 / 18.44 & 88.97 & 59.25 &  & 75.09 / 13.31 & 46.25 &  & 76.09 / 18.41 & 74.52 \\
Triplet & GloVe & Tag & 89.62 / 35.64 & 82.09 / 28.64 &  & 86.45 / 18.38 & 88.62 & 58.13 &  & 73.91 / 12.64 & 48.75 &  & 75.73 / 17.87 & 73.69 \\
Triplet & GloVe & Sentence & 89.67 / 35.58 & 82.38 / 28.82 &  & 86.51 / 18.54 & \textbf{89.31} & 58.25 &  & 74.17 / 12.75 & 48.75 &  & 75.74 / 17.79 & 74.38 \\
Triplet & GloVe & Stochastic & 89.07 / 34.08 & 82.11 / 28.24 &  & 86.74 / 18.35 & \textbf{89.31} & 55.62 &  & 74.67 / 12.61 & 51.25 &  & 75.82 / 17.93 & 73.96 \\
Triplet & BERT$_{base}$ & Tag & 89.28 / 34.44 & 81.56 / 26.74 &  & 85.38 / 16.67 & 84.48 & 54.87 &  & 73.53 / 11.87 & 47.50 &  & 72.89 / 17.39 & 73.62 \\
Triplet & BERT$_{base}$ & Sentence & 89.63 / 35.12 & 82.13 / 28.02 &  & 86.24 / 17.81 & 88.62 & 57.75 &  & 74.71 / 12.79 & 48.75 &  & 75.2 / 17.48 & 74.19 \\
Triplet & BERT$_{base}$ & Stochastic & 89.45 / 34.60 & 81.95 / 27.91 &  & 86.20 / 18.00 & 86.90 & 57.75 &  & 74.35 / 12.19 & \textbf{52.50} &  & 74.01 / 17.24 & 74.08 \\
Contrastive & BERT$_{base}$ & Tag & 90.95 / 38.08 & \textbf{83.10 / 29.75} &  & \textbf{87.52 / 19.88} & \textbf{89.31} & 58.50 &  & 75.64 / 14.19 & 46.25 &  & \textbf{76.83 / 18.83} & \textbf{75.49} \\
Contrastive & BERT$_{base}$ & Sentence & 90.34 / 37.39 & 82.29 / 27.95 &  & 86.34 / 17.64 & 85.17 & 57.75 &  & 74.77 / 13.11 & 48.75 &  & 73.72 / 17.4 & 74.70 \\
Contrastive & BERT$_{base}$ & Stochastic & \textbf{91.11 / 38.37} & 82.87 / 29.74 &  & 87.50 / 19.57 & 88.97 & \textbf{60.00} &  & \textbf{76.25 / 13.95} & 48.75 &  & 76.65 / 18.98 & 75.31 \\
\rowcolor[HTML]{EFEFEF}  
\multicolumn{3}{l}{\color[HTML]{656565}State-of-the-art \cite{huang2022mulan,alonso2022music,matthew2022supervised,alonso2022music,manco2022learning,matthew2022supervised,alonso2022music,kim2021learning}} & \color[HTML]{656565}{92.7 / -} & \color[HTML]{656565}{84.3 / 32.1} &  & \color[HTML]{656565}{87.7 / 20.3} & \color[HTML]{656565}{83.5} & \color[HTML]{656565}{61.1} &  & \color[HTML]{656565}{78.6 / 16.1} & - &  & \color[HTML]{656565}{78.8 / 20.2} & \color[HTML]{656565}{74.7} \\ \bottomrule
\end{tabular}
}
\caption{Zero-shot Transfer and Probing Evaluation.}
\label{tab:evaluation}
\end{table*}

\subsection{Evaluation}
\noindent \textbf{Text-based Retrieval} Depending on the type of input query, text-based music retrieval is divided into tag-level and sentence-level. Since the evaluation of tag-level retrieval is the same as label-wise evaluation of the auto-tagging task, we use the conventional macro version of ROCAUC and PRAUC metrics \cite{choi2019zero, won2020multimodal}. We report both evaluation results on the top 50 vocabularies of CALS \cite{won2021transformer} and the 1054 large vocabularies of ECALS\footnote{Since ECALS includes all tags of CALS, both cases can be evaluated with one ECALS pre-trained model}. For the evaluation of sentence-level retrieval, we build an audio-sentence subset by randomly sampling 1000 (audio, sentence) pairs from our testing split. Following the previous work \cite{manco2022learning}, the sentence-level retrieval performance is evaluated by measuring Recall at K (K={1,5,10}), mean average Precision at 10 (mAP10), and Median Rank (MedR). In case of the classification model, we annotate multi-label tags on the music items with the best f1 score thresholds. And we perform sentence-level retrieval on the frequency of words overlapping with the sentence query.

\vspace{3mm}
\noindent \textbf{Zero-shot Transfer and Probing} For evaluation of unseen query retrieval and generalization ability, we measure the zero-shot transfer and probing performance, respectively. The zero-shot transfer measures the prediction score as the cosine similarity between the audio embedding of music and the text embedding of unseen tag \cite{choi2019transfer}. For the probing task, we trained two shallow classifiers (linear models and one-layer MLPs) with the average pooled embedding from the frozen audio encoder. For rigorous comparison, we follow the probing protocol of previous studies \cite{manco2022learning,castellon2021codified}.

\subsection{Training Details}
The input to the audio encoder is a 9.91-second audio signal at 16~kHz sampling rate. It is converted to a log-scaled mel~spectrogram with 128 mel bins, 1024-point FFT with a hann window, and a hop size of 10~ms. During training, we randomly sample audio chunk from 30 seconds of the waveform. All models are optimized using Adam and use a 64-batch size. We use different learning rates for text encoders. The models that do not use the text encoder (classification and triplet-GloVe) were trained with a learning rate of 1e-3. The models with the BERT text encoder were with a learning rate of 5e-5. Contrastive-loss models were trained with a 0.2 temperature $\tau$, and triplet-loss models are with a 0.4 margin $\delta$.


\section{Results}
Table \ref{tab:text_retireval} shows the retrieval performances of different models using tag-level and sentence-level inputs. Firstly, the classification model is a competitive baseline for tag-based retrieval (Table~\ref{tab:text_retireval}-left). Although the model cannot generalize to unseen tags (even if they are synonyms or acronyms), the classification model is a reliable solution when abundant music tags are available for training. However, the classification model could not handle sentence-level inputs because it's only trained with tag-level queries due to its inherent design.

The pre-trained language model is versatile enough to handle both tag-level and sentence-level inputs. The pre-trained word embedding could also take sentence-level inputs by averaging the word embeddings, but the performance is not comparable. One possible reason is that the language model can summarize the sequence better than simple averaging. Another possible explanation is that the language model (BERT) was trained with larger data than the word embedding. Our proposed stochastic sampling approach further improves the performance when it's applied to the text encoder.

Contrastive learning consistently showed better retrieval performance than triplet approaches in tag-level and sentence-level inputs, although we used elaborated negative sampling. We interpret that larger negative sampling from a batch is more suitable than triplet sampling in retrieval tasks. In summary, contrastive learning of text-music representation using a pre-trained language model and stochastic sampling achieved the best retrieval performance.

We report the zero-shot transfer and probing results in Table~\ref{tab:evaluation}. Similar to the retrieval task, the contrastive-loss model in the zero-shot transfer task showed robust performance in almost all datasets. Compared to recent text-music representation learning approaches \cite{choi2019transfer, manco2022contrastive, huang2022mulan}, we see that contrastive-loss models achieve competitive results and show significant improvements on the MTAT and GTZAN dataset. All probing results of contrastive-loss models are close to the state-of-the-art performance and achieve state-of-the-art performance on GTZAN and KVT datasets. 

We also believe the inclusion of large-scale data can improve the performance. Recent multimodal representation learning approaches \cite{radford2021learning} have shown breakthough in many domains by taking advantage of enormous data from the web. A similar trend is found in our downstream evaluation. Contrastive learning of text-music representation using 44 million data \cite{huang2022mulan} significantly outperforms other approaches trained with 0.5 to 3.3 million dataset \cite{castellon2021codified, alonso2022music, matthew2022supervised} in MTAT tagging. Unfortunately, MTAT was the only common dataset with reported performance across various previous works.


\section{Conclusion}
In this paper, we introduced effective design choices for universal text-to-music retrieval. Recent text-music representation learning frameworks are assessed by using a carefully designed dataset and downstream tasks. We mainly focused on training objectives and text representation. Experimental results revealed that retrieval performance heavily depends on text representation. And contrastive models achieve better performance than triplet models in both retrieval and downstream tasks. Furthermore, our proposed stochastic text representation achieved robust performance in tag-level, caption-level, and zero-shot query retrieval cases. However, our current dataset is limited to music tags, such as genre, mood, and instrument. A more generalizable music retrieval system needs to cover other musical attributes, such as the tempo, key, chord progression, melody, artist, etc. To overcome the limitations of annotated labels, multi-task learning of multiple datasets or a teacher-student model can be an alternative. Reproducible code, pre-trained models~\footnote{https://github.com/seungheondoh/music-text-representation}, dataset~\footnote{https://github.com/seungheondoh/msd-subsets} and the proposed benchmark~\footnote{https://github.com/seungheondoh/msu-benchmark} are available online for future research.

\bibliographystyle{IEEEbib}
\bibliography{strings,refs}
\vfill\pagebreak
\appendix
\section{Supplement Material}
In this section, we provide additional details of the presented experiments including a list of the datasets, models, and qualitative and quantitative results. 

\subsection{ECALS Dataxset}
We perform tag- and sentence-level retrieval evaluation using the ECALS (Extended Clean tag and Artist-Level Stratified) dataset. As introduced in~\ref{sec:ecals}, the ECALS dataset is proposed based on the Million Song Dataset(MSD)~\cite{bertin2011million}. The MSD is a collection of metadata and audio features of 1 million tracks. The extended Last.fm annotation provides tags of more than 500,000~songs with 522,366 distinct tags whose distribution follows a long tail distribution. There are multiple MSD subsets depending on the post processing. The top50s subset~\cite{choi2016automatic, nam2018deep} consists of top-50 popular tags including genre, mood, instrument, vocal, and decade. Later, the CALS~\cite{won2021transformer} subset was proposed to solve the artist information leakage problem of the top50s subset. However, due to the small vocabulary, the two subsets were not suitable for caption-level text representation. To address this problem, we propose ECALS (Extended Clean tag and Artist-Level Stratified) subset. ECALS was created by merging the CALS subset with 500 Last.fm tags \cite{won2020multimodal} and 1,402 AllMusic \cite{schindler2019multi} tag annotation. The ECALS subset has 0.52 million 30-second clips and 140k unique tag captions. The advantage of merging multiple datasets is covering larger multiple tag categories including genre, style, instrument, vocal, mood, theme, and culture. 


\begin{table*}[!t]
\centering
\resizebox{0.95 \textwidth}{!}{%
\begin{tabular}{llllllccc}
\toprule
 &  &  &  &  &  & MTAT$^{\flat}$ & MTAT$^{\sharp}$ & GZ \\
Model & Loss & Audio Enc. & Text Enc. & Text Rep. & Pretrain-Dataset & ROC/PR & ROC/PR & ACC \\ \midrule
\multicolumn{9}{l}{{\color[HTML]{656565} \textit{Zeroshot Transfer:}}} \\
Choi et al \cite{choi2019zero} & Triplet & 1D CNN & GloVe & Tag & MSD$_{\text{ZSL}}$ \text{(0.41M)} & - & 73.9 / - - - & 73.1 \\
MusCALL$_{\text{base}}$ \cite{manco2022contrastive} & Contrastive & ResNet & SenBERT & Sentence & Production Music (0.25M) & - & 78.0 / 28.3 & 55.5 \\
MusCALL$_{\text{SSL}}$  \cite{manco2022contrastive} & Contrastive & ResNet & SenBERT & Sentence & Production Music (0.25M) & - & 77.4 / \textbf{29.3} & 58.2 \\
MuLan \cite{huang2022mulan} & Contrastive & ResNet & BERT$_{base}$ & Sentence & Youtube Music (44M) & 78.2 / - - - & - & - \\
\textbf{Ours} & Contrastive & Transformer & BERT$_{base}$ & Stochastic & MSD$_\text{ECALS}$ (0.52M) & \textbf{78.4} / 21.2 & \textbf{78.7} / 25.2 & \textbf{87.9} \\ \midrule
\multicolumn{8}{l}{{\color[HTML]{656565} \textit{Probing:}}} \\
MuLaP \cite{manco2022learning} & Alignment, MM & Musicnn & BERT$_{base}$ & Sentence & Production Music (0.25M) & - & 89.3 / 40.2 & - \\
MuLan \cite{huang2022mulan} & Contrastive & ResNet & BERT$_{base}$ & Sentence & Youtube Music (44M) & \textbf{92.7} / - - - & - & - \\
\textbf{Ours} & Contrastive & Transformer & BERT$_{base}$ & Stochastic & MSD$_\text{ECALS}$ (0.52M) & 91.1 / 38.4 & \textbf{91.7} / \textbf{46.1} & \textbf{89.0} \\ \bottomrule
\end{tabular}
}
\caption{Comparisons to the state-of-the-art of Music-Langauge Representations. \textbf{MM} stands for intra-modality Masked Modelling.}
\label{tab:text-music}
\end{table*}

\begin{table*}[!t]
\centering
\resizebox{\textwidth}{!}{%
\begin{tabular}{llcccccccccccc}
\toprule
&  & \multicolumn{2}{c}{Tagging} &  & \multicolumn{3}{c}{Genre} &  & \multicolumn{2}{c}{Mood/Theme} &  & \multicolumn{2}{c}{Inst/vocal} \\
 &  & MTAT$^{\flat}$ & MTG-top50s &  & MTG-G & GZ & FMA &  & MTG-M & Emoti &  & MTG-I & KVT \\
Model Type & Pretrain-Dataset & ROC/PR & ROC/PR &  & ROC/PR & ACC & ACC &  & ROC/PR & ACC &  & ROC/PR & F1 \\ \midrule

\multicolumn{14}{l}{{\color[HTML]{656565} \textit{Baseline:}}} \\
VGGish \cite{alonso2022music, hershey2017cnn} & YouTube Video (8M) &  90.2 / 37.2 & 83.2 / 28.2 &  & 86.3 / 17.2 & - & 53.0 &  & 76.3 / 14.1 & - &  & \textbf{78.8 / 20.2} & - \\ 
\midrule
\multicolumn{14}{l}{{\color[HTML]{656565} \textit{Audio-Music Representation Learning:}}} \\
CALM \cite{castellon2021codified} & OpenAI (1.2M) &  91.5 / 41.4  & - &  & - & 79.7 & - &  & - & - &  & - & - \\ 
Discog-Artist \cite{alonso2022music} & Discog (3.3M) &  90.7 / 38.0 & 83.6 / 30.6 &  & \textbf{87.7 / 20.3} & - & 59.1 &  & 76.3 / 14.3 & - &  & 69.7 / 16.9 & - \\
Musicset-Sup \cite{matthew2022supervised} & Musicset (1.8M) &  91.7 / 41.3  & \textbf{84.3 / 32.1} &  & - & 83.5 & - &  & \textbf{78.6 / 16.1} & - &  & - & - \\ \midrule
\multicolumn{14}{l}{{\color[HTML]{656565} \textit{Text-Music Representation Learning:}}} \\
MuLap \cite{manco2022learning} & Production Music (0.25M) & - & 82.6 / 27.3 &  & 85.9 / - & - & \textbf{61.1} &  & 76.1 / - & - &  & 76.8 / - & - \\
MuLan \cite{huang2022mulan} & Youtube Music (44M) & \textbf{92.7} / - & - &  & - & - & - &  & - & - &  & - & - \\ 
\textbf{Ours} & MSD$_\text{ECALS}$ (0.52M) & 91.1 / 38.4 & 82.9 / 29.7 &  & 87.5 / 19.6 & \textbf{89.0} & 60.0 &  & 76.3 / 14.0 & 48.8 &  & 76.7 / 19.0 & 75.3 \\
\bottomrule
\end{tabular}
}
\caption{Comparisons to the state-of-the-art \textbf{Probing} performance of Music Representation Learnings.}
\label{tab:allrep}
\end{table*}

\subsection{Downstream Dataset}
For downstream evaluation, we use multiple heterogeneous datasets. Each dataset has been actively used for assessing generalizability of pretrained models \cite{alonso2022music, matthew2022supervised}. 

\vspace{3mm}
\noindent\textbf{MagnaTagATune (MTAT)} \cite{law2009evaluation} consists of 26k music clips from 5,223 unique songs. This dataset has two splits\footnote{\text{https://github.com/jordipons/musicnn-training/tree/master/data/index/mtt}}\footnote{\text{https://github.com/jongpillee/music\_dataset\_split/tree/master/MTAT\_split}} depending on the inclusion or deletion of audio segments without any associated tags. In Table~\ref{tab:downstream}, we choose one
split as previous works \cite{castellon2021codified, huang2022mulan}. However, if only one split is used, comparison with previous works \cite{manco2022contrastive, manco2022learning} is impossible. To solve this problem, we compare our frameworks on both splits in Table~\ref{tab:text-music}. 

\vspace{3mm}
\noindent\textbf{MTG-Jamendo (MTG)} \cite{bogdanov2019mtg} contains 55,000 full-length audio tracks with 195 tags from the Jamendo music platform. All tags are annotated by categories such as genre, instrument, and mood/theme. We use an official split (\textit{split-0}) in each category.

\vspace{3mm}
\noindent\textbf{GTZAN (GZ)} \cite{tzanetakis2002musical} contains 30-second clips from 10 distinct genres for a single-label multi-class classification. We employ the fault-filtered version of this dataset \cite{sturm2013gtzan} \footnote{https://github.com/jongpillee/music\_dataset\_split/tree/master/GTZAN\_split}.

\vspace{3mm}
\noindent\textbf{FMA (FMA-Small)} \cite{defferrard2016fma}  is a large-scale public dataset with a rich set of metadata (e.g, tag, artist, ,.etc), user data, audio, and features. For evaluation, we use the Small subset of FMA, which contains 8,000 tracks and 8 genre tags: \textit{Hip-Hop, Pop, Folk, Experimental, Rock, International, Electronic}, and \textit{Instrumental}.

\vspace{3mm}
\noindent\textbf{Emotify (Emoti)} \cite{aljanaki2016studying} consists of 400 music excerpts in 4 genres such as rock, classical, pop, and electronic. The annotations were collected using Geneva Emotional Music Scales (GEMS)\footnote{http://www2.projects.science.uu.nl/memotion/emotifydata}. We use categorical nine emotion words:  \textit{Amazement, Solemnity, Tenderness, Nostalgia, Calmness, Power, Joyful, Tension}, and \textit{Sadness}.

\vspace{3mm}
\noindent\textbf{K-pop Vocal Tag (KVT)} \cite{kim2020semantic} consists of 6,787 vocal segments from K-pop music tracks\footnote{https://khlukekim.github.io/kvtdataset/}. They are annotated with 42 semantic tags which describe various vocal characteristics in the categories of pitch range, timbre, playing techniques, and gender.


\subsection{Comparison of Text-Music Representation Learning}
Table \ref{tab:text-music} shows the zero-shot transfer and probing performance of state-of-the-art (SoTA) text-music representations. For a clear comparison, we use a different split (MTAT$^{\sharp}$) \cite{won2020evaluation, manco2022contrastive, manco2022learning} as well as the split used in the paper (MATA$^{\flat}$) \cite{huang2022mulan, matthew2022supervised, alonso2022music}. As mentioned before, our approach shows SoTA performance on both MTAT$^{\flat}$ and GTZAN datasets except PRAUC score on MTAT$^{\sharp}$. We believe that this is due to the difference in the text representation. The models trained with the MSD dataset (Ours and \cite{choi2019zero}) perform better on multi-class genre classification (GTZAN) regardless of modality encoder and training objectives. This indicates that text representation is a critical element of the text-music representation learning framework. In the case of MTAT$^{\sharp}$, which is a multi-label task, the model \cite{manco2022learning} trained with expert annotation caption of production music shows a higher PRAUC score (25.2$\xrightarrow{}$29.3). Compared to the model trained on the 44M youtube dataset \cite{huang2022mulan}, our proposed model slightly outperforms on zero-shot transfer performance but performs significantly worse in the probing task. This is because the quality and size of the dataset are critical for the high-level music semantic task.

\begin{figure}[!t]
\centering
\includegraphics[width=\columnwidth]{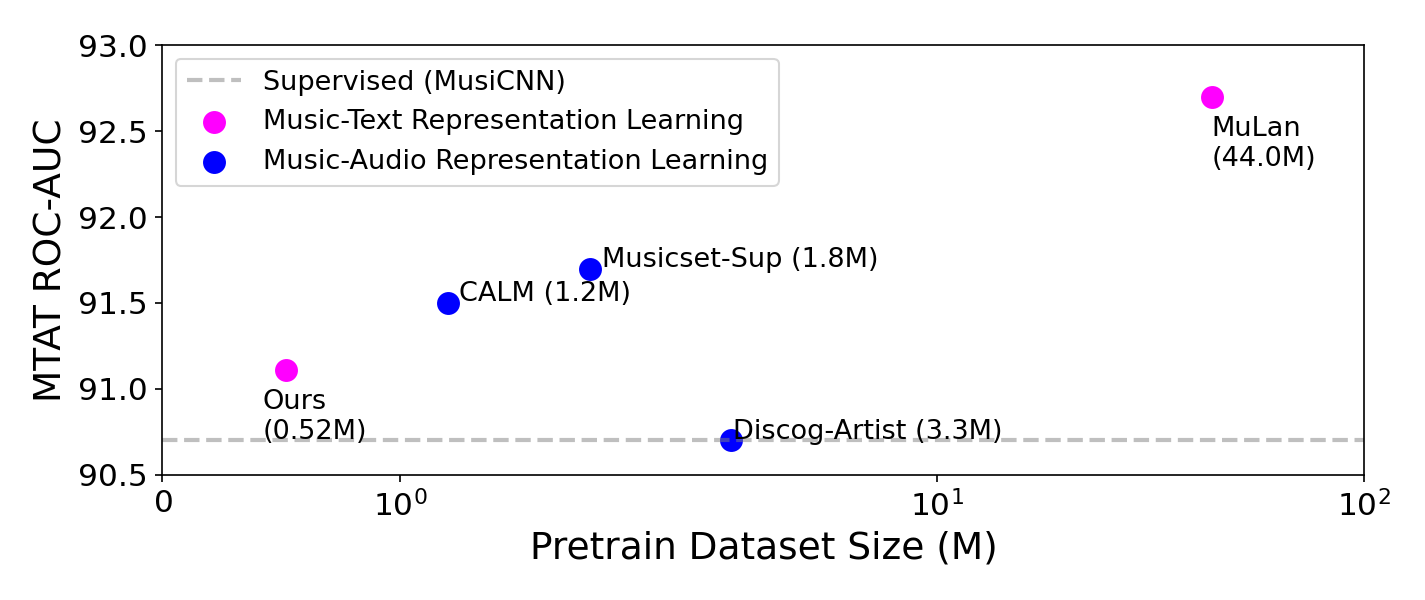}
\caption{MTAT ROC-AUC performance of state-of-the-arts model}
\label{fig:com}
\end{figure}

\begin{table*}[!t]
\resizebox{\textwidth}{!}{%
\begin{tabular}{lllcccccccccccc}
\toprule
 &  &  & \multicolumn{2}{c}{Tagging} &  & \multicolumn{3}{c}{Genre} &  & \multicolumn{2}{c}{Mood/Theme} &  & \multicolumn{2}{c}{Inst/vocal} \\
 &  &  & MATA$^{\flat}$ & MTG-top50s &  & MTG-G & GZ & FMA &  & MTG-M & Emoti &  & MTG-I & KVT \\
Model Type & Text Enc. & Text Rep. & ROC/PR & ROC/PR &  & ROC/PR & ACC & ACC &  & ROC/PR & ACC &  & ROC/PR & F1 \\ \midrule
\multicolumn{15}{l}{{\color[HTML]{656565} \textit{Linear Probing:}}} \\
Classification & Binary & Tag & 88.35 / 33.81 & 82.13 / 27.63 &  & 86.28 / 17.41 & 88.28 & 58.63 &  & 74.06 / 12.76 & 46.25 &  & 75.42 / 18.15 & 74.38 \\
Triplet & GloVe & Tag & 88.07 / 33.83 & 81.32 / 27.42 &  & 85.85 / 17.53 & 88.97 & 56.75 &  & 73.08 / 11.98 & 51.25 &  & 73.15 / 16.78 & 73.64 \\
Triplet & GloVe & Sentence & 88.28 / 33.66 & 81.83 / 27.80 &  & 86.21 / 17.81 & 88.97 & 56.62 &  & 73.59 / 12.40 & 52.50 &  & 74.63 / 17.23 & 74.21 \\
Triplet & GloVe & Stochastic & 87.37 / 32.22 & 81.84 / 27.36 &  & 86.20 / 17.65 & 89.31 & 54.75 &  & 73.92 / 11.62 & 51.25 &  & 73.49 / 16.84 & 73.91 \\
Triplet & BERT$_{base}$ & Tag & 87.55 / 32.00 & 80.96 / 25.55 &  & 84.32 / 15.52 & 82.07 & 52.62 &  & 72.59 / 11.31 & 46.25 &  & 72.84 / 15.97 & 73.60 \\
Triplet & BERT$_{base}$ & Sentence & 88.50 / 34.02 & 81.72 / 26.92 &  & 85.75 / 17.22 & 88.97 & 54.75 &  & 74.06 / 12.34 & 50.00 &  & 74.45 / 17.13 & 74.07 \\
Triplet & BERT$_{base}$ & Stochastic & 87.89 / 33.42 & 81.51 / 26.68 &  & 85.60 / 17.11 & 88.28 & 56.38 &  & 73.57 / 11.80 & 48.75 &  & 72.83 / 16.81 & 74.26 \\
Contrastive & BERT$_{base}$ & Tag & 90.18 / 36.99 & 82.58 / 28.93 &  & 86.86 / 18.84 & 88.97 & 57.38 &  & 74.74 / 13.50 & 50.00 &  & 76.72 / 19.14 & 75.27 \\
Contrastive & BERT$_{base}$ & Sentence & 89.23 / 35.16 & 81.74 / 26.81 &  & 85.47 / 16.73 & 86.21 & 54.37 &  & 74.07 / 12.56 & 48.75 &  & 73.50 / 17.43 & 74.61 \\
Contrastive & BERT$_{base}$ & Stochastic & 90.35 / 37.37 & 82.71 / 28.74 &  & 86.60 / 18.69 & 89.31 & 56.50 &  & 75.17 / 13.41 & 47.50 &  & 74.89 / 18.71 & 75.13 \\ \midrule
\multicolumn{15}{l}{{\color[HTML]{656565} \textit{MLP Probing:}}} \\
Classification & Binary & Tag & 89.72 / 35.54 & 82.66 / 28.78 &  & 87.01 / 18.44 & 88.97 & 59.25 &  & 75.09 / 13.31 & 46.25 &  & 76.09 / 18.41 & 74.52 \\
Triplet & GloVe & Tag & 89.62 / 35.64 & 82.09 / 28.64 &  & 86.45 / 18.38 & 88.62 & 58.13 &  & 73.91 / 12.64 & 48.75 &  & 75.73 / 17.87 & 73.69 \\
Triplet & GloVe & Sentence & 89.67 / 35.58 & 82.38 / 28.82 &  & 86.51 / 18.54 & 89.31 & 58.25 &  & 74.17 / 12.75 & 48.75 &  & 75.74 / 17.79 & 74.38 \\
Triplet & GloVe & Stochastic & 89.07 / 34.08 & 82.11 / 28.24 &  & 86.74 / 18.35 & 89.31 & 55.62 &  & 74.67 / 12.61 & 51.25 &  & 75.82 / 17.93 & 73.96 \\
Triplet & BERT$_{base}$ & Tag & 89.28 / 34.44 & 81.56 / 26.74 &  & 85.38 / 16.67 & 84.48 & 54.87 &  & 73.53 / 11.87 & 47.50 &  & 72.89 / 17.39 & 73.62 \\
Triplet & BERT$_{base}$ & Sentence & 89.63 / 35.12 & 82.13 / 28.02 &  & 86.24 / 17.81 & 88.62 & 57.75 &  & 74.71 / 12.79 & 48.75 &  & 75.20 / 17.48 & 74.19 \\
Triplet & BERT$_{base}$ & Stochastic & 89.45 / 34.60 & 81.95 / 27.91 &  & 86.20 / 18.00 & 86.90 & 57.75 &  & 74.35 / 12.19 & 52.50 &  & 74.01 / 17.24 & 74.08 \\
Contrastive & BERT$_{base}$ & Tag & 90.95 / 38.08 & 83.10 / 29.75 &  & 87.52 / 19.88 & 89.31 & 58.50 &  & 75.64 / 14.19 & 46.25 &  & 76.83 / 18.83 & 75.49 \\
Contrastive & BERT$_{base}$ & Sentence & 90.34 / 37.39 & 82.29 / 27.95 &  & 86.34 / 17.64 & 85.17 & 57.75 &  & 74.77 / 13.11 & 48.75 &  & 73.72 / 17.40 & 74.70 \\
Contrastive & BERT$_{base}$ & Stochastic & 91.11 / 38.37 & 82.87 / 29.74 &  & 87.50 / 19.57 & 88.97 & 60.00 &  & 76.25 / 13.95 & 48.75 &  & 76.65 / 18.98 & 75.31 \\
\bottomrule
\end{tabular}
}
\caption{Linear and MLP Probing Evaluation}
\label{tab:linmlp}
\end{table*}

\subsection{Comparison of State-of-the-art-models}
Table \ref{tab:allrep} shows the probing performance of various music representations. Following the previous work \cite{alonso2022music}, we select the pre-trained VGGish model \cite{hershey2017cnn} as a baseline. The large-scale music domain dataset outperformed the baseline in high-level semantic tasks (general tagging, genre, mood) regardless of the training framework. The baseline model performed better in the relatively low-level semantic task (instrument). It is expected since the datasets used in pretraining (MSD, Discog, Production Music) consist of only multi-track recordings while YouTube videos contain both single- and multi-track instrument data.

In Figure~\ref{fig:com}, we show the MTAT$^{\flat}$ ROC-AUC results of different music representation learning models. Our 0.5M ECALS pre-trained model outperforms the supervised MusiCNN \cite{pons2018atscale} baseline and the 3.3M Discog pre-trained model \cite{alonso2022music}. This shows that our proposed approach is efficient with the relatively small size of pre-trained dataset. MuLan \cite{huang2022mulan}, a text-music representation model with 44M Youtube music dataset, demonstrates impressive performance with a huge gap. It refers to the importance of the dataset size. However, the comparisons against these frameworks may be unreliable due to the differences in their training datasets, modality encoders, and data representations.

\subsection{Comparison between Linear and MLP Probing}
In the probing task \cite{van2014transfer, alain2016understanding}, we take the audio features from the frozen encoder and fit a linear or multi-layer perceptron (MLP) classifier to predict the target classes. In Table~\ref{tab:linmlp}, we report both classifier performance for \textit{linear} and \textit{non-linear} separability of audio features. Across all the categories of music semantics, MLP classifiers outperform linear classifiers. It is interpreted that \textit{non-linearity} is more suitable for the music semantic understanding task, presumably because the music is multi-label data with higher-level semantic labels.

\begin{figure}[!t]
\centering
\includegraphics[width=\columnwidth]{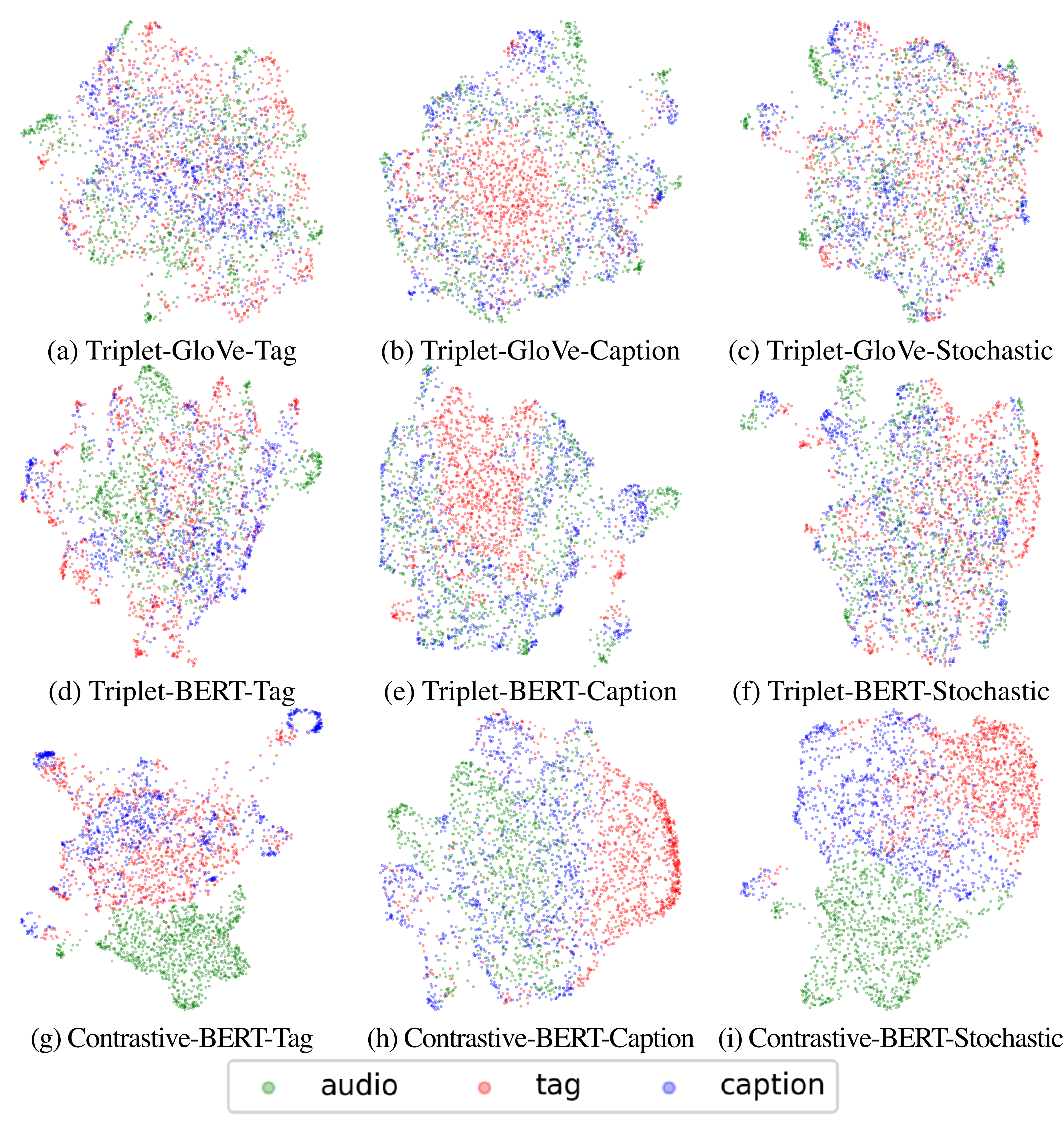}
\caption{UMAP visualization of audio-tag-caption joint embedding space. \textbf{[First Row]} Triplet framework with Glove word encoder. \textbf{[Second Row]} Triplet framework with BERT word encoder. \textbf{[Third Row]} Contrastive framework with BERT word encoder. Each column shows tag, caption, and stochastic text representation respectively.}
\label{fig:umap}
\end{figure}

\subsection{Visualization}

The multimodal embedding spaces are projected to a 2D space using uniform manifold approximation and projection (UMAP) \cite{mcinnes2018umap}. We fit UMAP with music-audio, caption, and tag embeddings and then projected all embeddings (In Figure~\ref{fig:umap}). For the dataset, ECALS 1000 audio-caption pairs and 1054 tags were used. The contrastive model shows a more significant gap between audio and text modality than the triplet models. However, it is difficult to find the correlation between the visualized distribution and the performance reported above (Table~\ref{tab:text_retireval}, Table~\ref{tab:evaluation}, Table~\ref{tab:linmlp}), Compared to the triplet model, the stochastic model shows a more entangled embedding space than other tag and caption models. In the second column, the caption-based model, it is interesting that the tag embeddings are isolated. This supports the example in the table above where caption models showed low performance in tag-based retrieval tasks. Contrary to this, caption and tag embeddings are mixed up in the tag-based model.

\end{document}